\documentclass{ws-procs9x6}
\def\1{\mbox{l\hspace{-0.53em}1}}

\begin{document}

\title{Excited Baryons in the $1/N_c$ Expansion}

\author{N. Matagne and Fl. Stancu}

\address{University of Li\`ege, Physics Department,\\
Institute of Physics, B.5, \\
Sart Tilman, B-4000 Li\`ege 1, Belgium\\
E-mail: nmatagne@ulg.ac.be,
fstancu@ulg.ac.be}

\maketitle

\abstracts{
We review results for the mass spectrum of orbitally excited baryons obtained in the $1/N_c$ expansion. We show the dependence of various contributions to the mass operator as a function of the excitation energy.}

\section{Introduction}
At low energies, typical for baryon spectroscopy, QCD does not admit a perturbative expansion in the strong coupling constant. About 30 years ago, 't Hooft suggested an alternative approach based on an $1/N_c$ expansion where $N_c$ is the number of colors\cite{HOOFT}. Witten described the counting rules for such an expansion\cite{WITTEN}. The method works very well for the ground state baryons inasmuch as they display an SU($2N_f$) exact symmetry, where $N_f$ is the number of flavors\cite{DM93}.
Although for excited states this symmetry is broken, in the last few years it has been realized that an $1/N_c$ expansion can as well be used  to describe states belonging to various SU(6) multiplets. A particular attention has been paid to the $[{\bf 70},1^-]$ multiplet\cite{Goi97}\cdash\cite{SGS}.

Here we review recent work on the mass spectrum of baryons in the $N=2$ and $N=4$ bands. We are especially concerned with orbitally excited baryons belonging to the $[{\bf 56},2^+]$, Ref. [\refcite{GSS03}], $[{\bf 70},\ell^+]\ (\ell=0,2)$, Ref. [\refcite{MS2}] and $[{\bf 56},4^+]$, Ref. [\refcite{MS1}], multiplets .

\section{Mass operator}

The mass operator up to $\mathcal{O}(1/N_c)$ has the general form
\begin{equation}
M = \sum c_iO_i + \sum b_i\bar{B}_i
\end{equation}
where the operators $O_i$ are SU(3)-flavor singlets and the operators $\bar{B}_i$, which are defined to have vanishing expectation values for non-strange states, break the flavor symmetry. The coefficients $c_i$ and $b_i$ that encode QCD dynamics are evaluated by a numerical fit to the available data. The operators $O_i$ and $\bar{B}_i$ can be expressed as positive parity and rotationally invariant products of generators of SU(6) $\otimes$ O(3).

The analysis of symmetric $[{\bf 56}, \ell^+]$ states is rather simple\cite{GSS03,MS1}. The total wave functions  are obtained by coupling an orbital part $\sim Y_{\ell m}$ to spin-flavor symmetric states.
But for  mixed-symmetric  representations, it is necessary to split the wave function into two parts : a symmetric core composed of $N_c-1$ quarks and a excited quark.
Generally, in the case of the mixed symmetric representations, one has both core generators $\ell_c^i, S_c^i, T^a_i$ and $G^{ia}_c$ and  excited quark generators $\ell_q^i, s^i, t^a$ and $g^{ia}$. The multiplet $[{\bf 70},1^-]$ is a particular case with $\ell_c = 0$.

As an illustration, Table \ref{operators} gives the list of operators chosen for the study of the $[{\bf 70},\ell^+]$ multiplets\cite{MS2}. $O_1$ is the SU(6) scalar operator of order $N_c$, $O_2$ and $O_4$ are the dominant parts of the spin-orbit and spin-spin operators respectively.  Note that $O_2$ is $\mathcal{O}(N_c^0)$ for mixed-symmetric states, in contrast to the symmetric states case where it is $\mathcal{O}(N_c^{-1})$, see Refs. [\refcite{MS2,GOITY04}]. $O_3 \sim \mathcal{O}(N_c^0)$ due to $G^{ja}_c$. Strange baryons are not included in this study. Table \ref{operators} contains the values of the coefficients $c_i$ obtained by fitting the available experimental data. We present in Table \ref{multiplet} the masses of the resonances which we have interpreted as belonging to the $[{\bf 70},0^+]$ or $[{\bf 70},2^+]$ multiplets.

\begin{table}
\tbl{List of operators and the coefficients resulting from the fit with $\chi^2_{\rm dof}  \simeq 0.83$ for the $[{\bf 70},\ell^+]$ multiplets.}
{\normalsize
\label{operators}
\renewcommand{\arraystretch}{1.2}
\begin{tabular}{llrrl}
\hline
Operator & \multicolumn{4}{c}{Fitted coef. (MeV)}\\
\hline
$O_1 = N_c \ \1 $                                   & \ \ \ $c_1 =  $  & 555 & $\pm$ & 11       \\
$O_2 = \ell_q^i s^i$                                & \ \ \ $c_2 =  $  &   47 & $\pm$ & 100    \\
$O_3 = \frac{3}{N_c}\ell^{(2)ij}_{q}g^{ia}G_c^{ja}$ & \ \ \ $c_3 =  $   & -191 & $\pm$ & 132  \\
$O_4 = \frac{1}{N_c}(S_c^iS_c^i+s^iS_c^i)$          & \ \ \ $c_4 =  $  &  261 & $\pm$ &  47  \\
\hline
\end{tabular}}
\end{table}

\begin{table}
\tbl{The partial contribution and the total mass (MeV) predicted for the $[{\bf 70},\ell^+]$ by the $1/N_c$ expansion as compared with the empirically known masses.}
{\tiny
\label{multiplet}
\renewcommand{\arraystretch}{1.5}
\begin{tabular}{lrrrrccl}\hline \hline
                    &      \multicolumn{5}{c}{$1/N_c$ expansion results}        &    &                     \\
\cline{2-6}
                    &      \multicolumn{4}{c}{Partial contribution} &  Total & Empirical  &   Name, status  \\
\cline{2-5}
                    &     $c_1O_1$  &  $c_2O_2$ & $c_3O_3$ & $c_4O_4$   &    &        \\
\hline
$^4N[{\bf 70},2^+]\frac{7}{2}^+$        & 1665 & 31 & 42 & 217 &      $ 1956\pm95$  & $2016\pm104$ &  $F_{17}(1990)$**  \\
$^2N[{\bf 70},2^+]\frac{5}{2}^+$      & 1665 & 10   & 0 & 43 &      $1719\pm34 $  &    \\

$^4N[{\bf 70},2^+]\frac{5}{2}^+$   & 1665 & -5  & -106 &  217 &     $ 1771\pm88$  & $1981\pm200$ & $F_{15}(2000)$**  \\
$^4N[{\bf 70},0^+]\frac{3}{2}^+$   & 1665  & 0   & 0 & 217 &    $1883\pm17$  & $1879\pm17$ &  $P_{13}(1900)$** \\
$^2N[{\bf 70},2^+]\frac{3}{2}^+$  &   1665   &   -16  & 0 &  43   & $1693\pm42$  &                                \\
$^4N[{\bf 70},2^+]\frac{3}{2}^+$     &  1665    &   -31  & 0 & 217    & $1851\pm69$  &                                   \\
$^2N[{\bf 70},0^+]\frac{1}{2}^+$   &   1665   &   0  &  0 & 43  &   $1709\pm25$  &     $1710\pm30$          &    $P_{11}(1710)$***                 \\
$^4N[{\bf 70},2^+]\frac{1}{2}^+$   & 1665  & -47   & 149 & 217 &     $1985\pm26 $  &   $1986\pm26$ &  $P_{11}(2100)$*\\
\hline
$^2\Delta[{\bf 70},2^+]\frac{5}{2}^+$  &  1665    &   -10  &  0 & 87  &    $1742\pm29$  &  $1976\pm237$             &  $P_{35}(2000)$**                 \\
$^2\Delta[{\bf 70},2^+]\frac{3}{2}^+$     &   1665   &  16   &  0 & 87    &  $1768\pm38$  &                                   \\
$^2\Delta[{\bf 70},0^+]\frac{1}{2}^+$   &   1665   & 0    &  0 &  87 &   $1752\pm19$  &   $1744\pm36$            &   $P_{31}(1750)$*                  \\
\hline
\hline
\end{tabular}}
\end{table}

\section{The dependence of the coefficients $c_i$ on the excitation energy}

\begin{figure}[h]
\begin{center}
\includegraphics[width=5cm]{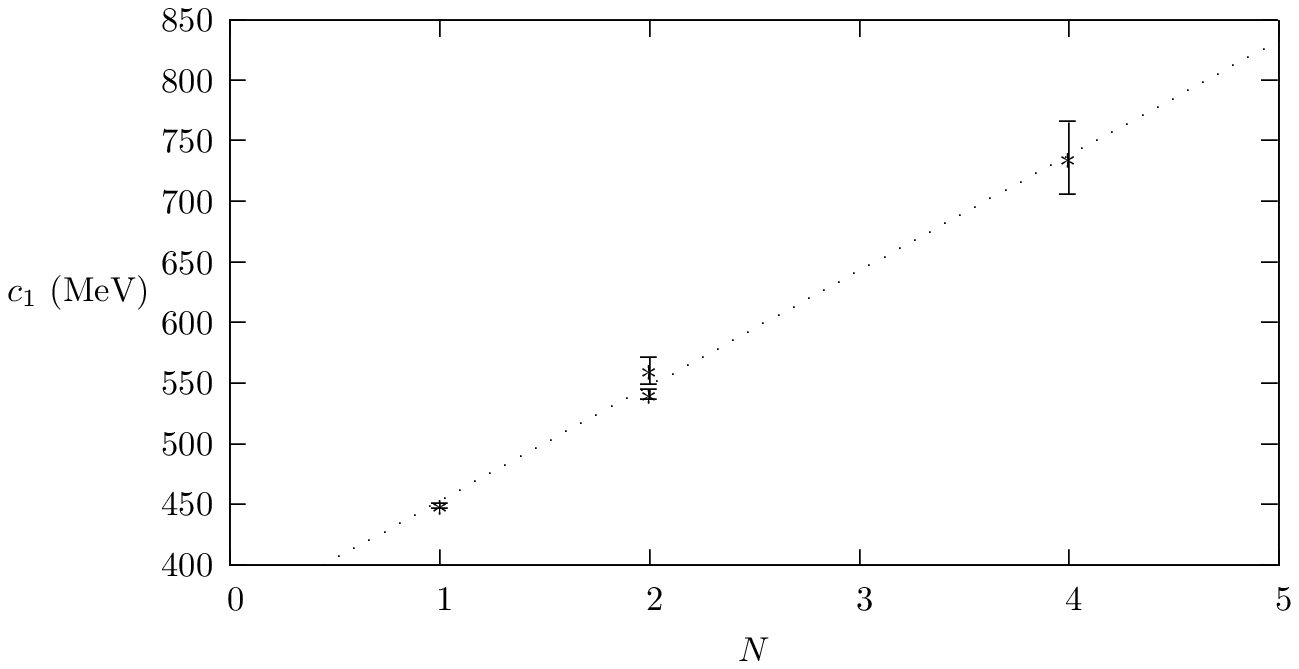}
\hspace{0.3cm}
\includegraphics[width=5cm]{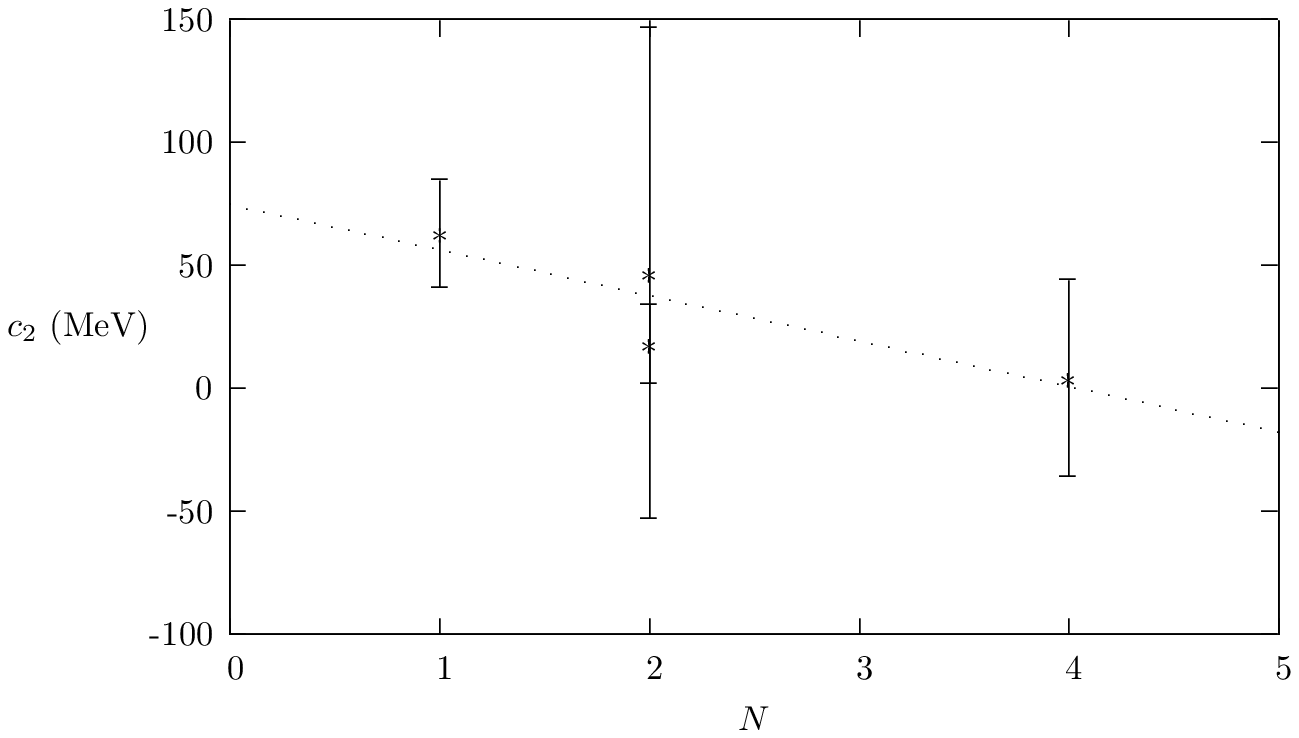} \\
\vspace{0.5cm}
\includegraphics[width=5cm]{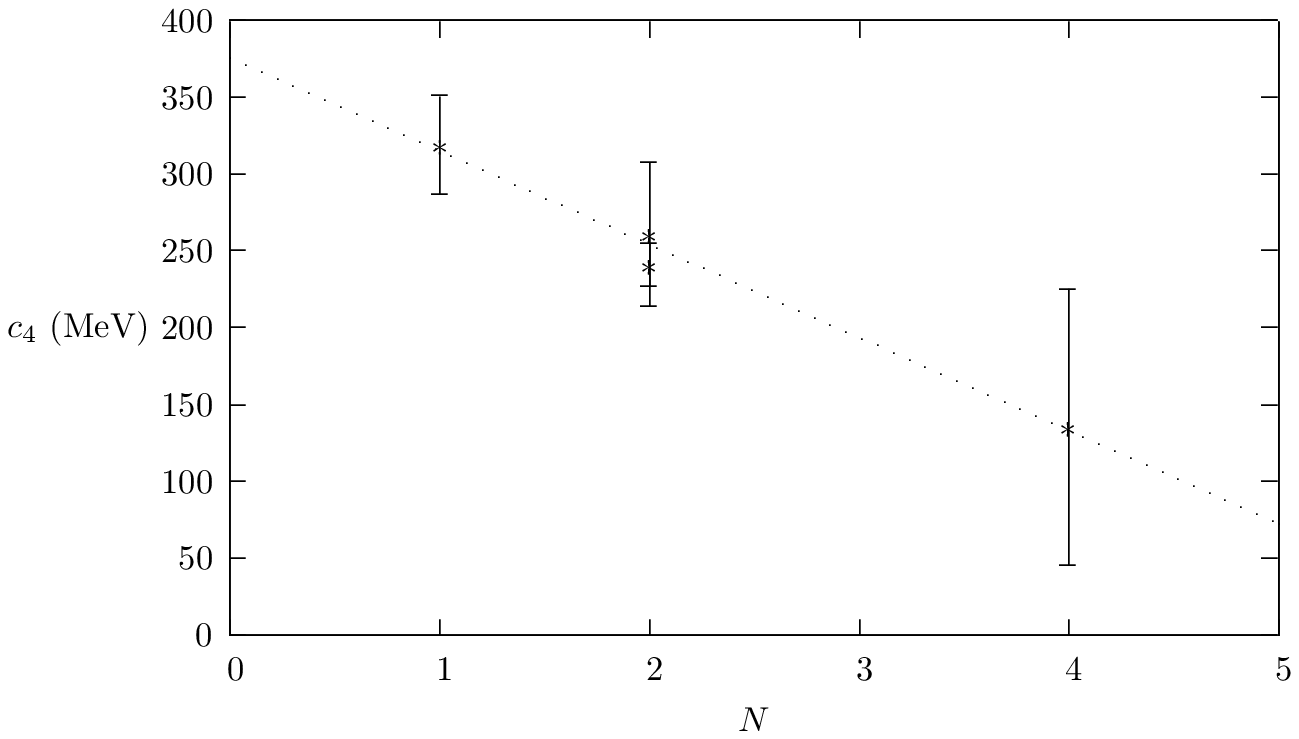}
\caption{Evolution of the coefficients $c_i$ with the excitation energy corresponding to $N=1,2$ and 4. The straight lines are to guide the eye.}
\label{evol}
\end{center}
\end{figure}
It is interesting to see the change of $c_i$ with the excitation energy.
In Figure \ref{evol}, we collect the presently know values of $c_i$ ($i=1,2,4$) with error bars for the orbitally excited states studied so far in the large $N_c$ expansion: $N=1$, Ref. [\refcite{SGS}], $N=2$ (lower values\cite{GSS03}, upper values\cite{MS2}) and $N=4$, Ref. [\refcite{MS1}].
This behavior shows that at large energies the dominant contribution comes from $c_1$ and the contributions of the spin-dependent terms vanish. These results are consistent with the quark model picture where the linear term in $N_c$ contains the free mass term, the kinetic and the confinement energy. An intuitive model based on the chiral symmetry restoration has already predicted that the spin dependent interactions vanish at high energies\cite{LYG}.

\section{Conclusions}
Our work is based on the assumption that there is no multiplet mixing and is restricted to non-strange baryons. Future work is devoted to strange baryons. To better understand the applications of the $1/N_c$ expansion more and better data is desirable.

\section*{Acknowledgments}
The work of one of us (N. M.) was supported by the Institut Interuniversitaire des Sciences Nucl\'eraires (Belgium).

\end{document}